\documentclass[fleqn,12pt,twoside]{article}
\usepackage{espcrc1}
\usepackage{graphicx}

\newcommand\be{\begin{equation}}
\newcommand\ee{\end{equation}}
\newcommand\bea{\begin{eqnarray}}
\newcommand\eea{\end{eqnarray}}
\newcommand\gsim {~^{>~}_{\sim~}}
\newcommand\lsim {~^{<~}_{\sim~}}

\title{Quark Confinement Physics from Quantum Chromodynamics}

\author{H.~Suganuma$^{\rm a}$, K.~Amemiya\address{Research Center for Nuclear Physics (RCNP), Osaka University \\ 
Ibaraki, Osaka 567-0047, Japan}, H.~Ichie\address{Department of Physics, Tokyo Institute of Technology \\ 
Ohokayama 2-12-1, Meguro, Tokyo 152-8551, Japan} and A. Tanaka$^{\rm a}$}

\begin{document}
\maketitle

\begin{abstract} 
We show the construction of the dual superconducting theory 
for the confinement mechanism from QCD in the maximally abelian 
(MA) gauge using the lattice QCD Monte Carlo simulation. 
We find that essence of infrared abelian dominance is naturally 
understood with the off-diagonal gluon mass 
$m_{\rm off} \simeq 1.2 {\rm GeV}$ induced by the MA gauge fixing.
In the MA gauge, the off-diagonal gluon amplitude is forced to be small, 
and the off-diagonal gluon phase tends to be random. 
As the mathematical origin of abelian dominance for confinement, 
we demonstrate that the strong randomness of the off-diagonal gluon 
phase leads to abelian dominance for the string tension. 
In the MA gauge, there appears the macroscopic network of 
the monopole world-line covering the whole system. 
We investigate the monopole-current system in the MA gauge 
by analyzing the dual gluon field $B_\mu$. 
We evaluate the dual gluon mass as $m_B = 0.4 \sim$ 0.5GeV 
in the infrared region, which is the lattice-QCD evidence of 
the dual Higgs mechanism by monopole condensation. 
Owing to infrared abelian dominance and infrared monopole condensation, 
QCD in the MA gauge is describable with the dual Ginzburg-Landau theory. 
\end{abstract} 

\section{QCD and Dual Superconducting Theory for Confinement}

Since 1974, quantum chromodynamics (QCD) has been established 
as the fundamental theory of the strong interaction, 
however, it is still hard to understand 
the nonperturbative QCD (NP-QCD) phenomena such as 
color confinement and dynamical chiral-symmetry breaking, 
in spite of the simple form of the QCD lagrangian 
\bea
{\cal L}_{\rm QCD}=-{1 \over 2} {\rm tr} G_{\mu\nu}G^{\mu\nu}
+\bar q (i\gamma_\mu D^\mu-m_q) q. 
\eea
In particular, to understand the confinement mechanism is 
one of the most difficult problems remaining in the particle physics.
As the hadron Regge trajectory and the lattice QCD simulation show, 
the confinement force between the color-electric charges is 
characterized by 
the {\it one-dimensional squeezing} of the color-electric flux 
and the universal physical quantity of the {\it string tension} 
$\sigma \simeq 1{\rm GeV/fm}$.

As for the confinement mechanism, 
Nambu first proposed the {\it dual superconducting theory} for quark 
confinement, based on the electro-magnetic duality in 1974.$^{1}$ 
In this theory, there occurs the one-dimensional squeezing 
of the color-electric flux between quarks 
by the dual Meissner effect 
due to condensation of bosonic color-magnetic monopoles.
However, there are {\it two large gaps} between QCD and the 
dual superconducting theory.$^{2}$

\ 

\noindent
\begin{minipage}{.5cm}
\baselineskip .55cm
(1)~\\~\\
\end{minipage}
\begin{minipage}{15.3cm}
The dual superconducting theory is based on the 
{\it abelian gauge theory} subject to the Maxwell-type equations, 
where electro-magnetic duality is manifest, 
while QCD is a nonabelian gauge theory. 
\end{minipage}

\ 

\noindent
\begin{minipage}{.5cm}
\baselineskip .6cm
(2)~\\~\\
\end{minipage}
\begin{minipage}{15.3cm}
The dual superconducting theory requires condensation of 
color-magnetic monopoles as the key concept, while QCD does not 
have such a monopole as the elementary degrees of freedom.
\end{minipage}

\noindent
These gaps can be simultaneously fulfilled by 
the use of the {\it MA gauge fixing,} 
which reduces QCD to an abelian gauge theory. 
In the MA gauge, the off-diagonal gluon 
behaves as a charged matter field 
similar to $W^\pm_\mu$ in the Standard Model and provides 
a color-electric current in terms of the residual abelian gauge 
symmetry.
As a remarkable fact in the MA gauge, 
color-magnetic monopoles appear as topological objects 
reflecting the nontrivial homotopy group 
$\Pi_2($SU($N_c$)/U(1)$^{N_c-1})$
$=${\bf Z}$^{N_c-1}_\infty$, 
similarly in the GUT monopole.$^{3-6}$

Thus, in the MA gauge, QCD is reduced into an abelian gauge theory 
including both the electric current $j_\mu$ and 
the magnetic current $k_\mu$, which is expected 
to provide the theoretical basis of the dual superconducting theory 
for the confinement mechanism.

\section{MA Gauge Fixing and Extraction of 
Relevant Mode for Confinement}

In the Euclidean QCD, the maximally abelian (MA) gauge is 
defined by minimizing$^{2,6}$ 
\be 
R_{\rm off} [A_\mu ( \cdot )] \equiv \int d^4x {\rm tr}
[\hat D_\mu ,\vec H][\hat D_\mu ,\vec H]^\dagger
={e^2 \over 2} \int d^4x \sum_\alpha  |A_\mu ^\alpha (x)|^2, 
\ee
with the ${\rm SU}(N_c)$ covariant derivative operator 
$\hat D_\mu \equiv \hat \partial_\mu+ieA_\mu $ and 
the Cartan decomposition $A_\mu (x)=\vec A_\mu (x) \cdot \vec H 
+\sum_\alpha A_\mu^\alpha (x)E^\alpha $. 
{\it In the MA gauge, the off-diagonal gluon components 
are forced to be as small as possible by the $SU(N_c)$ 
gauge transformation.} 
Since the covariant derivative $\hat D_\mu$ obeys the 
adjoint gauge transformation, the local form of 
the MA gauge fixing condition is derived as $^{2,6}$ 
\be
[\vec H, [\hat D_\mu , [\hat D_\mu , \vec H]]]=0.
\ee
(For $N_c=2$, this condition is equivalent to the diagonalization of 
$\Phi_{\rm MA}\equiv [\hat{D_\mu}, [\hat{D_\mu},\tau^3]]$,
and then the MA gauge is found to be 
a sort of the 't Hooft abelian gauge$^{3}$.) 
In the MA gauge, the gauge symmetry 
$G \equiv {\rm SU}(N_c)_{\rm local}$ 
is reduced into $H \equiv {\rm U(1)}_{\rm local}^{N_c-1} 
\times {\rm Weyl}_{\rm global}^{N_c}$, 
where the global Weyl symmetry is the subgroup of ${\rm SU}(N_c)$ 
relating the permutation of the $N_c$ bases in the fundamental 
representation.$^{2,6}$ 

We summarize abelian dominance, monopole dominance and 
extraction of the relevant mode for NP-QCD 
observed in the lattice QCD in the MA gauge.

\ 

\noindent
\begin{minipage}{.5cm}
\baselineskip .3cm
(a)~\\~\\
\end{minipage}
\begin{minipage}{15.3cm}
Without gauge fixing, 
all the gluon components equally contribute to NP-QCD, and 
it is difficult to extract relevant degrees of freedom for NP-QCD. 
\end{minipage}

\noindent
\begin{minipage}{.5cm}
\baselineskip .53cm
(b)~\\~\\~\\~\\~\\~\\~\\
\end{minipage}
\begin{minipage}{15.3cm}
In the MA gauge, QCD is reduced into an abelian gauge theory 
including the electric current $j_\mu $ 
and the magnetic current $k_\mu $. 
The diagonal gluon behaves as the abelian gauge field, 
and the off-diagonal gluon behaves as the charged matter field.
In the MA gauge, the lattice QCD shows {\it abelian dominance} 
for NP-QCD (confinement$^{6,7}$, chiral symmetry breaking$^{8}$, 
instantons$^{9}$) : 
only the diagonal gluon is relevant for NP-QCD, 
while off-diagonal gluons do not contribute to NP-QCD. 
In the lattice QCD, 
there appears the {\it global network of the monopole world-line 
covering the whole system in the MA gauge.} (See Fig.3(a).) 
\end{minipage}

\noindent
\begin{minipage}{.5cm}
\baselineskip .55cm
(c) ~\\ ~\\ ~\\  
\end{minipage}
\begin{minipage}{15.3cm}
The diagonal gluon can be decomposed into the 
``photon part'' and the ``monopole part'', 
corresponding to the separation of $j_\mu$ and $k_\mu$. 
In the MA gauge, the lattice QCD shows {\it monopole dominance} 
for NP-QCD:
the monopole part ($k_\mu \ne 0$, $j_\mu=0$) leads to NP-QCD, 
while the photon part ($j_\mu \ne 0, k_\mu=0$) seems 
trivial like QED and does not contribute to NP-QCD. 
\end{minipage}

\noindent 
Thus, by taking the MA gauge, 
the relevant collective mode for NP-QCD can be extracted 
as the color-magnetic monopole.$^{2,10}$

\section{Essence of Abelian Dominance :
   Off-diagonal Gluon Mass in MA Gauge}

In this section, we study essence of abelian dominance 
for NP-QCD in the MA gauge in terms of the {\it generation of the 
effective mass $m_{\rm off}$ of the off-diagonal (charged) gluon 
by the MA gauge fixing}. 
In the SU(2) QCD partition functional, the mass generation 
of the off-diagonal gluon $A_\mu^{\pm} \equiv 
(A_\mu^1 \pm i A_\mu^2)/\sqrt{2}$ in the MA gauge is expressed as$^{2}$ 
\bea
Z_{\rm QCD}^{\rm MA} &=& \int DA_\mu \exp\{iS_{\rm QCD}[A_\mu ]\} 
\delta (\Phi _{\rm MA}^\pm [A_\mu ])\Delta _{\rm PF}[A_\mu ] \cr
&=& 
\int DA_\mu ^3 \exp\{iS_{\rm eff}[A_\mu ^3]\}
\int DA_\mu ^\pm \exp\{i\int d^4x \ 
m_{\rm off}^2 A_\mu ^+A^\mu _- \} {\cal F}[A_\mu ],
\eea
with 
$
\Phi_{\rm MA} \equiv [\hat D_\mu , [\hat D^\mu , \tau_3]], 
$
the Faddeev-Popov determinant $\Delta _{\rm FP}$, 
the abelian effective action $S_{\rm eff}[A_\mu ^3]$ 
and a smooth functional ${\cal F}[A_\mu]$. 

To investigate the off-diagonal gluon mass $m_{\rm off}$, 
we study the Euclidean gluon propagator 
$G_{\mu \nu }^{ab} (x-y) \equiv \langle A_\mu ^a(x)A_\nu ^b(y)\rangle$ 
in the MA gauge, using the SU(2) lattice QCD.$^{2}$ 
As for the residual U(1)$_3$ gauge symmetry, 
we impose the U(1)$_3$ Landau gauge fixing 
to extract most continuous gauge configuration 
under the MA gauge constraint 
and to compare with the continuum theory.
The continuum gluon field $A_\mu^a(x)$ is extracted 
 from the link variable as 
$U_\mu(s)={\rm exp}(iaeA_\mu^a(s) \frac{\tau^a}{2})$.
Here, the scalar-type gluon propagator 
$G_{\mu\mu}^a(r)\equiv \sum^4_{\mu=1}\langle 
A_\mu^{~a}(x)A_\mu^{~a}(y)\rangle$ 
is useful to observe the interaction range of the gluon,
because it depends only on the four-dimensional Euclidean 
distance $r \equiv \sqrt{(x_\mu- y_\mu)^2}$.

We show in Fig.1(a) $G_{\mu \mu}^3(r)$ and 
$
G_{\mu\mu}^{+-}(r) \equiv 
\sum^4_{\mu=1}\langle A_\mu^{+}(x)A_\mu^{-}(y)\rangle
={1 \over 2} \{G_{\mu\mu}^1(r)+G_{\mu\mu}^2(r)\}
$
in the MA gauge using the SU(2) lattice QCD 
with $2.2 \le \beta \le 2.4$ and the various lattice size 
($12^3 \times 24$, $16^4$, $20^4$). 
Since the massive vector-boson propagator with the mass $M$ 
takes a Yukawa-type asymptotic form as $G_{\mu\mu}(r) \sim 
{M^{1/2} \over r^{3/2}}\exp(-Mr)$, 
the effective mass $m_{\rm off}$ of the off-diagonal gluon 
$A_\mu^{\pm}(x)$ can be evaluated from the slope of the logarithmic 
plot of $r^{3/2} G_{\mu\mu}^{+-}(r)\sim \exp(-m_{\rm off}r)$ 
as shown in Fig.1(b). 
The off-diagonal gluon $A_\mu^{\pm}(x)$ 
behaves as the massive field with $m_{\rm off} \simeq 1.2~{\rm GeV}$ 
in the MA gauge for $r \gsim 0.2$ fm. 

We perform also the standard mass measurement for the off-diagonal 
gluon in the similar manner to the hadron mass measurement 
in the lattice QCD. We calculate 
$
\Phi_\mu^\pm(\tau) \equiv \int d \vec x A_\mu^\pm(\vec x,\tau), 
$
and measure the temporal correlation of 
$
\Gamma_{\mu\mu}^{+-}(\tau) \equiv 
\langle \Phi_\mu^+(\tau)\Phi_\mu^-(0) \rangle 
$
in the MA gauge with the ${\rm U(1)}_3$ Landau gauge.
We obtain the off-diagonal gluon mass $m_{\rm off} \simeq 1.2 {\rm GeV}$ 
again from the slope of the logarithmic plot of 
$\Gamma_{\mu\mu}^{+-}(\tau)$ as the function of the 
temporal distance $\tau$ in the lattice QCD with 
$2.3 \le \beta \le 2.35$ with $16^3\times 32$ and $12^3\times 24$, 
as shown in Fig.1(c). 

Thus, {\it essence of infrared abelian dominance} 
in the MA gauge can be physically interpreted with 
the {\it effective off-diagonal gluon mass} 
$m_{\rm off}$ induced by the MA gauge fixing.$^{2}$ 
Due to the effective mass $m_{\rm off}\simeq 1.2 {\rm GeV}$, 
the off-diagonal gluon $A_\mu^\pm$ can propagate only within 
the short range as $r \lsim m_{\rm off}^{-1} \simeq 0.2 {\rm fm}$, 
and cannot contribute to the infrared QCD physics in the MA gauge, 
which leads to abelian dominance for NP-QCD.$^{2}$

\begin{figure}
\begin{center}
\includegraphics[scale=0.355]{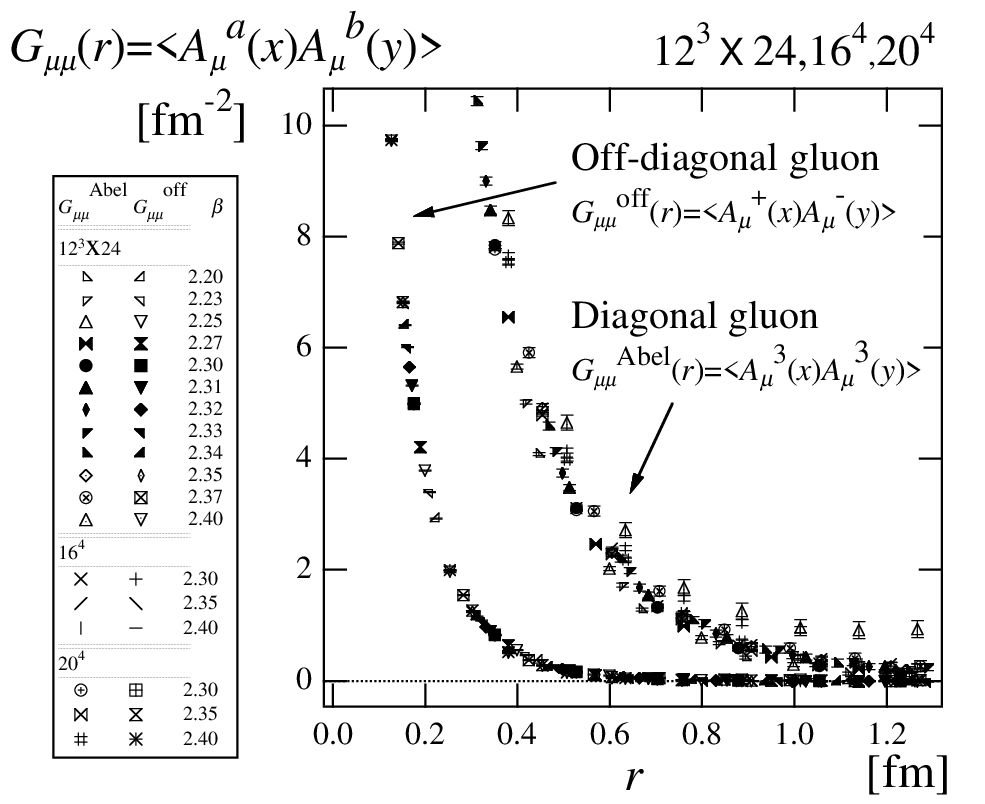}
\includegraphics[scale=0.38]{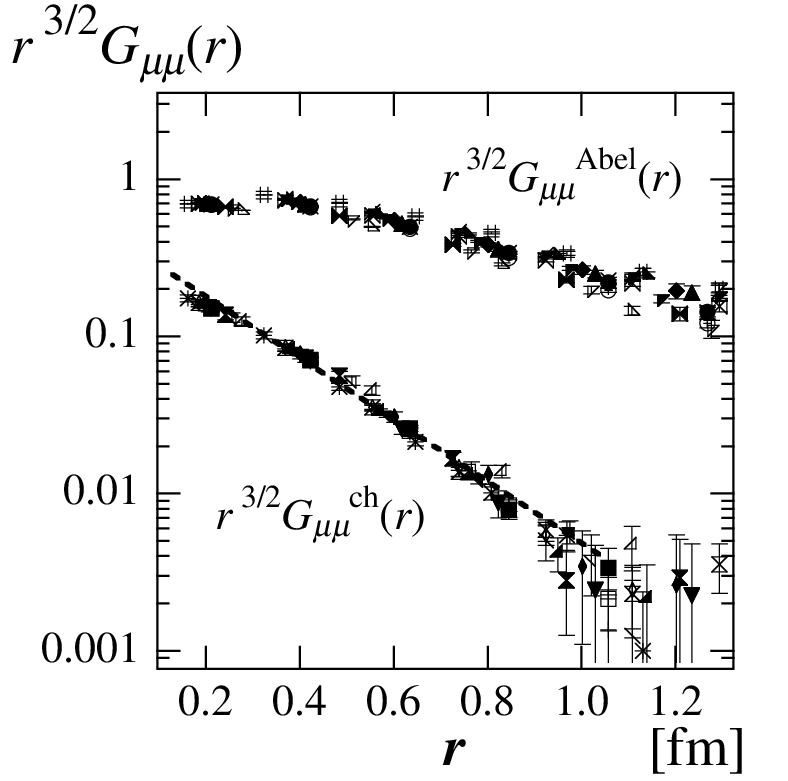}
\includegraphics[scale=0.405]{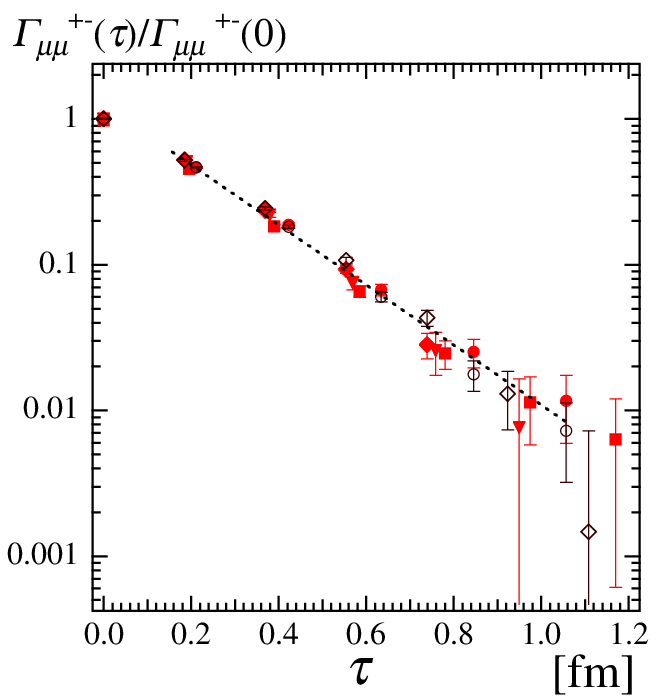}
\vspace{-1cm}
\caption{
(a) The scalar-type gluon propagator 
$G_{\mu \mu }^a(r)$ as the function of the 4-dimensional 
distance $r$ in the MA gauge in the SU(2) lattice QCD with 
$2.2 \le \beta \le 2.4$ and various lattice size 
($12^3 \times 24$, $16^4$, $20^4$). 
(b) The logarithmic plot for 
the scalar correlation $r^{3/2} G_{\mu \mu}^a(r)$. 
The off-diagonal gluon propagator behaves as 
the Yukawa-type function, 
$G_{\mu \mu } \sim {\exp(-m_{\rm off}r) \over r^{3/2}}$. 
(c) The logarithmic plot of the temporal correlation 
$\Gamma_{\mu\mu}^{+-}(\tau) \equiv 
\langle \Phi_\mu^+(\tau)\Phi_\mu^-(0) \rangle$
as the function of the temporal distance $\tau$ 
in the lattice QCD with $2.3 \le \beta \le 2.35$ 
with $16^3 \times 32$ and $12^3 \times 24$. 
From the slope of the dotted lines in (b) and (c), 
the effective mass of the off-diagonal gluon $A_\mu^\pm$ 
can be estimated as $m_{\rm off}\simeq 1.2 {\rm GeV}$.}
\end{center}
\vspace{-0.5cm}
\end{figure}

\section{Randomness of Off-diagonal Gluon Phase as 
the Mathematical Origin of Abelian Dominance for Confinement}

In the lattice QCD, the SU(2) link variable is factorized 
as $U_\mu(s)=M_\mu(s)u_\mu(s)$, according to the Cartan decomposition 
${\rm SU(2)/U(1)}_3 \times {\rm U(1)}_3$. 
Here, $u_\mu(s)\equiv \exp\{i\tau^3\theta^3_\mu(s)\}\in {\rm U(1)}_3$ 
denotes the abelian link variable, and the off-diagonal factor 
$M_\mu(s)\in {\rm SU(2)/U(1)}_3$ is parameterized as 
\be
M_\mu(s)\equiv e^{i\{\tau^1\theta^1_\mu(s)+\tau^2\theta^2_\mu(s)\}}
=\left( {\matrix{
{\rm cos}{\theta_\mu}(s) & -{\rm sin}{\theta_\mu}(s) e^{-i\chi_\mu(s)} \cr
{\rm sin}{\theta_\mu}(s) e^{i\chi_\mu(s)} & {\rm cos}{\theta_\mu}(s)
}} \right). 
\ee

In the MA gauge, the {\it diagonal element} $\cos \theta_\mu(s)$ 
in $M_\mu(s)$ is maximized by the SU(2) gauge transformation 
and the ``abelian projection rate'' becomes almost unity as 
$R_{\rm Abel}=\langle\cos \theta_\mu(s)\rangle_{\rm MA}\simeq 0.93$ 
at $\beta=2.4$. 
Using the lattice QCD simulation, 
we find the two remarkable features of the off-diagonal element 
$e^{i\chi_\mu(s)}\sin\theta_\mu(s)$ in $M_\mu(s)$ 
in the MA gauge.$^{2,6}$

\ 

\noindent
\begin{minipage}{.5cm}
(1)~\\
\end{minipage}
\begin{minipage}{15.3cm}
The off-diagonal gluon amplitude $|\sin\theta_\mu(s)|$ 
is forced to be minimized in the MA gauge, which allows 
the approximate treatment on the off-diagonal gluon phase. 
\end{minipage}
\noindent
\begin{minipage}{.5cm}
(2)~\\~\\
\end{minipage}
\begin{minipage}{15.3cm}
The off-diagonal phase variable $\chi_\mu(s)$ is not 
constrained by the MA gauge-fixing condition at all, 
and tends to be random.
\end{minipage}

\noindent
Therefore, $\chi_\mu(s)$ can be regarded as a 
{\it random angle variable} 
on the treatment of $M_\mu(s)$ in the MA gauge 
in a good approximation. 

Now, we show the analytical proof of abelian dominance 
for the string tension or the confinement force in the MA gauge, 
within the random-variable approximation for $\chi_\mu(s)$ 
or the off-diagonal gluon phase.$^{2,6}$
Here, we use
$
\langle e^{i\chi_\mu(s)}\rangle_{\rm MA} 
\simeq \int_0^{2\pi} d\chi_\mu(s)\exp\{i\chi_\mu(s)\}=0. 
$
In calculating the Wilson loop $\langle W_C[U]\rangle \equiv 
\langle{\rm tr}\Pi_C U_\mu(s) \rangle= 
\langle{\rm tr}\Pi_C\{M_\mu(s)u_\mu(s)\}\rangle$, 
the off-diagonal matrix $M_\mu(s)$ 
is simply reduced as a $c$-number factor, 
$
M_\mu(s) \rightarrow \cos \theta_\mu(s) \ {\bf 1},
$
and then the SU(2) link variable $U_\mu(s)$ 
is reduced to be a {\it diagonal matrix,} 
\be
U_\mu(s)\equiv M_\mu(s)u_\mu(s)
\rightarrow 
\cos \theta_\mu(s) u_\mu(s), 
\ee
after the integration over $\chi_\mu(s)$. 
For the $R \times T$ rectangular $C$, the Wilson loop 
$W_C[U]$ in the MA gauge is approximated as 
\bea
\langle W_C[U]\rangle 
	&=&
	\langle{\rm tr}\Pi_{i=1}^L 
	\{M_{\mu_i}(s_i)u_{\mu_i}(s_i)\}\rangle
	\simeq 
	\langle\Pi_{i=1}^L \cos \theta_{\mu_i}(s_i) \cdot 
	{\rm tr} \Pi_{j=1}^L u_{\mu_j}(s_j)\rangle_{\rm MA} \nonumber\\
	&\simeq&
	\langle\exp\{\Sigma_{i=1}^L 
		\ln (\cos \theta_{\mu_i}(s_i))\}\rangle_{\rm MA} 
	\ \langle W_C[u]\rangle_{\rm MA},
\eea
with the perimeter length $L \equiv 2(R+T)$ and 
the abelian Wilson loop 
$W_C[u] \equiv {\rm tr}\Pi_{i=1}^L u_{\mu_i}(s_i)$. 
Replacing $\sum_{i=1}^L \ln \{\cos(\theta_{\mu_i}(s_i))\}$ 
by its average 
$L \langle \ln \{\cos \theta_\mu(s)\} \rangle_{\rm MA}$
in a statistical sense, 
we derive a formula for the {\it off-diagonal gluon contribution 
to the Wilson loop} as$^{2,6}$ 
\bea
W_C^{\rm off}\equiv 
\langle W_C[U]\rangle/\langle W_C[u]\rangle_{\rm MA}
\simeq \exp\{L\langle \ln\{\cos \theta_\mu(s)\}\rangle_{\rm MA}\},
					\label{eqn:9}
\eea
which provides the relation between the {\it macroscopic} 
quantity $W_C^{\rm off}$ and the {\it microscopic} 
quantity $\langle \ln\{\cos \theta_\mu(s)\}\rangle_{\rm MA}$. 
Using the lattice QCD, we have checked this relation 
for large loops, where such a statistical treatment is accurate.$^{2,6}$ 

In this way, the off-diagonal gluon contribution $W_C^{\rm off}$ 
obeys the {\it perimeter law} in the MA gauge, 
and then the off-diagonal gluon contribution 
to the string tension vanishes as 
\bea
\sigma_{\rm SU(2)}-\sigma_{\rm Abel} 
\simeq 
-2 \langle \ln \{\cos\theta_\mu(s)\} \rangle_{\rm MA}
\lim_{R,T \rightarrow \infty} {R+T \over RT}=0.
\eea
Thus, {\it abelian dominance for the string tension}, 
$\sigma_{\rm SU(2)}=\sigma_{\rm Abel}$, 
can be demonstrated in the MA gauge within 
the random-variable approximation for the off-diagonal gluon phase. 
Also, we can predict the deviation between 
$\sigma_{\rm SU(2)}$ and $\sigma_{\rm Abel}$ as 
$\sigma_{\rm SU(2)} > \sigma_{\rm Abel}$, 
due to the {\it finite size effect} on $R$ and $T$ 
in the Wilson loop.$^{2,6}$ 

\section{The Structure of QCD-Monopoles in terms of the Off-diagonal Gluon}

Let us compare the QCD-monopole with the point-like Dirac monopole. 
There is no point-like monopole in QED, 
because the QED action diverges around the monopole. 
The QCD-monopole also accompanies a large abelian action density 
inevitably, however, {\it owing to cancellation with 
the off-diagonal gluon contribution, the total QCD action is 
kept finite even around the QCD-monopole.}$^{2,6}$ 

To see this, we investigate the structure of the QCD-monopole 
in the MA gauge in terms of the action density 
using the SU(2) lattice QCD.$^{2}$ 
From the SU(2) plaquette $P^{\rm SU(2)}_{\mu \nu }(s)$ 
and the abelian plaquette $P^{\rm Abel}_{\mu \nu }(s)$, 
we define the ``SU(2) action density'' 
$
S_{\mu \nu }^{\rm SU(2)}(s) 
\equiv 1-{1 \over 2}{\rm tr}P^{\rm SU(2)}_{\mu \nu }(s), 
$
the ``abelian action density'' 
$
S_{\mu \nu }^{\rm Abel}(s) 
\equiv 1-{1 \over 2}{\rm tr}P^{\rm Abel}_{\mu \nu }(s) 
$
and the ``off-diagonal gluon contribution'' 
$
S_{\mu \nu }^{\rm off}(s) 
\equiv S_{\mu \nu }^{\rm SU(2)}(s)-S_{\mu \nu }^{\rm Abel}(s). 
$
In the lattice formalism, 
the monopole current $k_\mu (s)$ is defined on the dual link, 
and there are 6 plaquettes around the monopole. 
Then, we consider the  
average over the 6 plaquettes around the dual link, 
\be
S(s,\mu) \equiv
{1 \over 12} \sum_{\alpha\beta\gamma} \sum_{m=0}^1 
| \varepsilon_{\mu\alpha\beta\gamma} |
S_{\alpha\beta}(s + m \hat \gamma).
\ee
We show in Fig.2(b) the probability distribution of 
the action densities 
$S_{\rm SU(2)}$, $S_{\rm Abel}$ and 
$S_{\rm off}$ around the QCD-monopole in the MA gauge. 
We summarize the results on the QCD-monopole structure as follows.\\ 

\noindent
\begin{minipage}{.5cm}
\baselineskip .39cm
(1)~\\ ~\\ ~\\ 
\end{minipage}
\begin{minipage}{15.3cm}
Around the QCD-monopole, 
both the abelian action density $S_{\rm Abel}$ 
and the off-diagonal gluon contribution $S_{\rm off}$ are 
largely fluctuated, and their cancellation keeps 
the total QCD-action density $S_{\rm SU(2)}$ small.
\end{minipage}
\begin{minipage}{.5cm}
\baselineskip .35cm
(2) ~\\  
\end{minipage}
\begin{minipage}{15.3cm}
The QCD-monopole has an intrinsic structure 
relating to a large amount of off-diagonal gluons $A_\mu^\pm$ 
around its center, 
similar to the 't~Hooft-Polyakov monopole. 
\end{minipage}
\begin{minipage}{.5cm}
\baselineskip .3cm
(3) ~\\  
\end{minipage}
\begin{minipage}{15.3cm}
At the large-distance scale, off-diagonal gluons 
inside the QCD-monopole become invisible, 
and the QCD-monopole can be regarded as the point-like 
Dirac monopole. 
\end{minipage}
\begin{minipage}{.5cm}
\baselineskip .55cm
(4) ~\\ ~\\ 
\end{minipage}
\begin{minipage}{15.3cm}
From the concentration of off-diagonal gluons 
around QCD-monopoles in the MA gauge, 
we can naturally understand the 
{\it local correlation between monopoles and instantons}. 
In fact, instantons tend to appear around the monopole world-line 
in the MA gauge, because instantons need 
full SU(2) gluon components for existence.$^{2,9}$ 
\end{minipage}

\begin{figure}
\begin{center}
\includegraphics[scale=0.4]{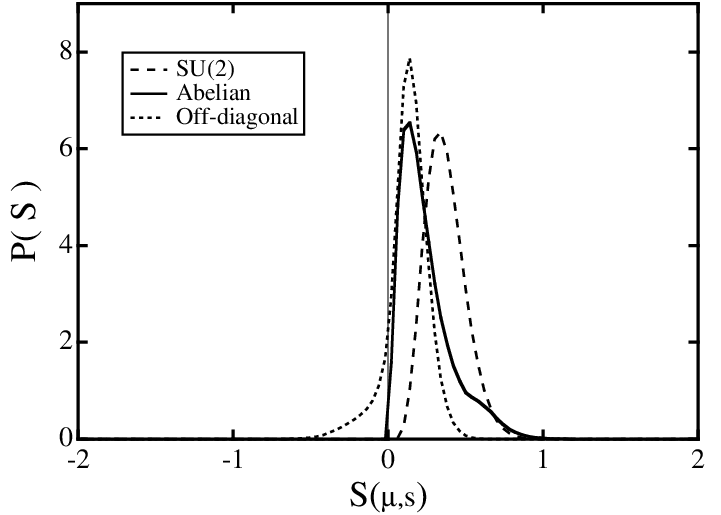}
\includegraphics[scale=0.4]{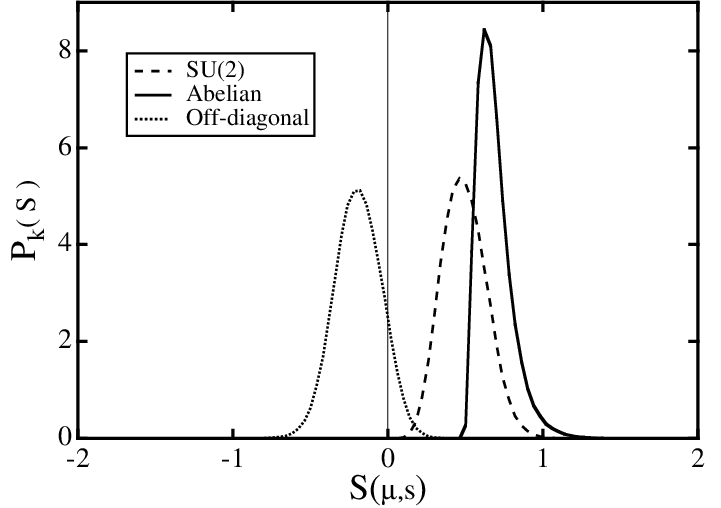}
\includegraphics[scale=0.42]{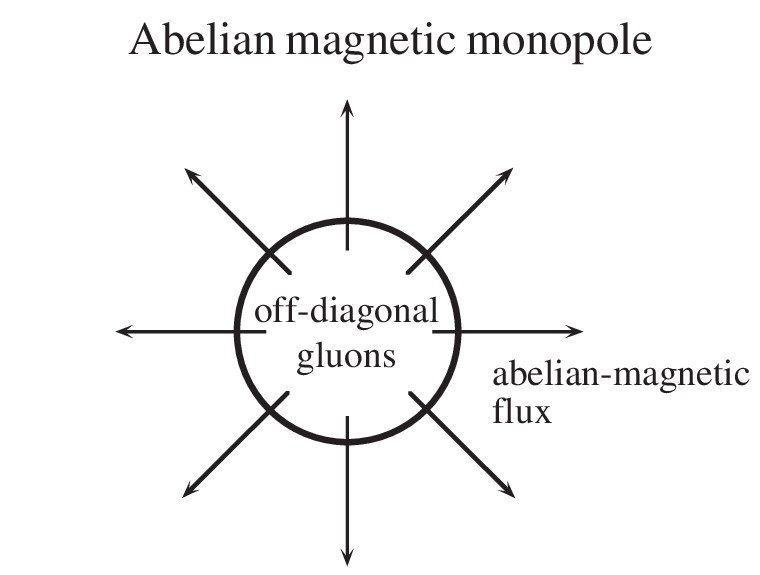}
\vspace{-0.5cm}
\caption{
(a) The total probability distribution $P(S)$ on the whole lattice 
and (b) the probability distribution $P_k(S)$ around the monopole 
for SU(2) action density $S_{\rm SU(2)}$ (dashed curve), 
abelian action density $S_{\rm Abel}$ (solid curve) 
and off-diagonal gluon contribution $S_{\rm off}$ (dotted curve) 
in the MA gauge at $\beta =2.4$ on $16^4$ lattice. 
Around the QCD-monopole, 
large cancellation between $S_{\rm Abel}$ and  
$S_{\rm off}$ keeps the total QCD-action small.  
(c) The schematic figure for the QCD-monopole structure 
in the MA gauge. 
The QCD-monopole includes a large amount of off-diagonal gluons 
around its center. }
\end{center}
\end{figure}

\section{Lattice-QCD Evidence of Infrared Monopole Condensation}

In the MA gauge, there appears the global network of the 
monopole world-line covering the whole system 
as shown in Fig.3(a), and 
this monopole-current system (the monopole part) 
holds essence of NP-QCD$^{2,8-10}$. 
We finally study the dual Higgs mechanism by 
monopole condensation in the NP-QCD vacuum in the MA gauge. 

Since QCD is described by the ``electric variable'' as 
quarks and gluons, 
the ``electric sector'' of QCD has been well studied with 
the Wilson loop or the inter-quark potential, however, 
the ``magnetic sector'' of QCD is hidden and still unclear. 
To investigate the magnetic sector directly,
it is useful to introduce the ``dual (magnetic) variable'' 
as the {\it dual gluon field} $B_\mu $, which is the dual partner 
of the diagonal gluon 
and directly couples with the magnetic current $k_\mu $.

Due to the absence of the electric current $j_\mu$ 
in the monopole part,
the dual gluon $B_\mu $ can be introduced 
as the regular field satisfying 
$(\partial \land B)_{\mu\nu}={^*\!F}_{\mu\nu}$ 
and the dual Bianchi identity, 
$
{\partial^{\mu}} {^*\!(}\partial \land B)_{\mu\nu}=j_\nu=0.
$
By taking the dual Landau gauge $\partial_\mu B^\mu=0$, 
the field equation is simplified as $\partial^2 B_\mu =k_\mu$, 
and therefore we obtain the dual gluon field $B_\mu$ 
from the monopole current $k_\mu$ as 
\bea
B_\mu (x) = ( \partial^{-2} k_\mu)(x)= -\frac{1}{4\pi^2} 
\int d^4y \frac{k_\mu(y)}{(x-y)^2}. 
\eea
In the monopole-condensed vacuum, 
the dual gluon $B_\mu $ is to be massive, 
and hence we investigate the dual gluon mass $m_B$ 
as the evidence of the dual Higgs mechanism. 

First, we put test magnetic charges in the monopole-current system 
in the MA gauge, and measure the inter-monopole potential $V_M(r)$ 
to get information about monopole condensation. 
Since the dual Higgs mechanism is the screening effect on the 
magnetic flux, the inter-monopole potential is expected to be 
short-range Yukawa-type. 
Using {\it the dual Wilson loop} $W_D$ as the loop-integral of 
the dual gluon, 
\bea
W_D(C) \equiv \exp\{i{e \over 2}\oint_C dx_\mu B^\mu \}=
\exp\{i{e \over 2}\int\!\!\!\int d\sigma_{\mu\nu}{^*\!F}^{\mu\nu}\},
\eea
the potential between the monopole and the anti-monopole 
can be derived as 
\bea
V_{M}(R) = -\lim_{T \rightarrow  \infty} {1 \over T}\ln 
\langle W_D(R,T) \rangle. 
\eea
Here, $W_D(C)$ is the {\it dual version of the abelian Wilson loop} 
$W_{\rm Abel}(C) \equiv \exp\{i{e \over 2}\oint_C dx_\mu A^\mu \}=
\exp\{i{e \over 2}\int\!\!\!\int d\sigma_{\mu\nu}{F}^{\mu\nu}\}$ 
and we have set the test monopole charge as $e/2$. 

We show in Fig.3(b) the inter-monopole potential $V_M(r)$ 
in the monopole part in the MA gauge.$^2$ 
Except for the short distance, the inter-monopole potential 
can be almost fitted by the Yukawa potential 
$V_M(r) = -{{(e/2)}^2 \over 4\pi}{e^{-m_Br} \over r}$, 
after removing the finite-size effect of the dual Wilson loop. 
In the MA gauge, the dual gluon mass is estimated as 
$m_B \simeq {\rm 0.5GeV}$ from the infrared behavior of $V_M(r)$. 

Second, we investigate also the scalar-type dual gluon propagator 
$\langle B_\mu(x)B_\mu(y)\rangle_{\rm MA}$ 
as shown in Fig.3(c), 
and estimate the dual gluon mass as $m_B \simeq 0.4$ GeV 
from its large-distance behavior. 

From these two tests, the dual gluon mass is evaluated as 
$m_B=0.4 \sim 0.5$ GeV, 
and this can be regarded as the lattice-QCD evidence 
for the dual Higgs mechanism 
by monopole condensation at the infrared scale.

\begin{figure}
\begin{center}
\includegraphics[scale=0.8]{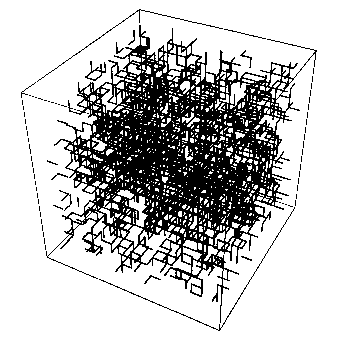}
\includegraphics[scale=0.35]{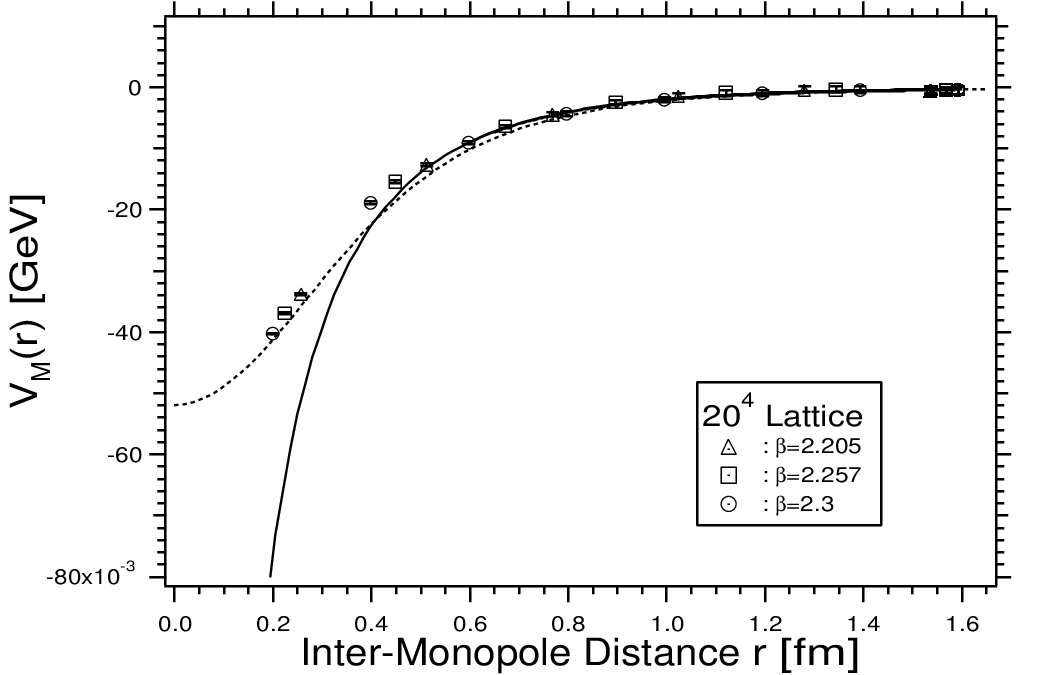}
\includegraphics[scale=0.35]{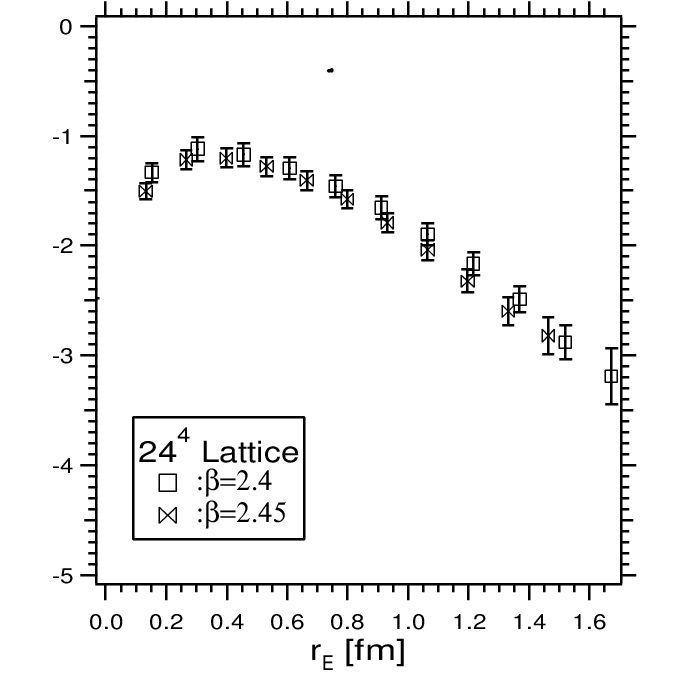}
\caption{
The SU(2) lattice-QCD results in the MA gauge. 
(a) The monopole world-line projected into ${\bf R}^3$ 
on the $16^3 \times 4$ lattice with $\beta =2.2$ 
(the confinement phase). 
There appears the global network of monopole currents covering 
the whole system. 
(b) The inter-monopole potential $V_M(r)$ 
v.s. the 3-dimensional distance $r$ 
in the monopole-current system on the $20^4$ lattice. 
The solid curve denotes the Yukawa potential with $m_B=0.5$GeV. 
The dotted curve denotes the Yukawa-type potential 
including the monopole-size effect. 
(c) The scalar-type dual-gluon correlation 
$\ln (r_E^{3/2} \langle B_\mu(x)B_\mu(y)\rangle_{\rm MA})$ as the 
function of the 4-dimensional Euclidean distance $r_E$ on the 
$24^4$ lattice.
}
\end{center}
\end{figure}

To summarize, the lattice QCD in the MA gauge exhibits 
{\it infrared abelian dominance} and 
{\it infrared monopole condensation}, 
and therefore the dual Ginzburg-Landau (DGL) theory 
can be constructed as the infrared effective theory 
directly based on QCD in the MA gauge.$^{2}$.

\end{document}